# PSNet: a deep learning model based digital phase-shifting algorithm from a single fringe image

Zhaoshuai Qi,* Xiaojun Liu, Xiaolin Liu, Jiaqi Yang, and Yanning Zhang

[1]National Engineering Laboratory for Integrated Aero-Space-Ground-Ocean Big Data Application Technology, College of Computer Science, Northwestern Polytechnical University, Xi'an 710072, China

*Corresponding author: zhaoshuaiqi1206@163.com

**As the gold standard for phase retrieval, phase-shifting algorithm (PS) has been widely used in optical interferometry, fringe projection profilometry, etc. However, capturing multiple fringe patterns in PS limits the algorithm to only a narrow range of application. To this end, a deep learning (DL) model based digital PS algorithm from only a single fringe image is proposed. By training on a simulated dataset of PS fringe patterns, the learnt model, denoted PSNet, can predict fringe patterns with other PS steps when given a pattern with the first PS step. Simulation and experiment results demonstrate the PSNet's promising performance on accurate prediction of digital PS patterns, and robustness to complex scenarios such as surfaces with varying curvature and reflectance. © 2021 Optical Society of America**

http://dx.doi.org/10.1364/OL.99.099999

In optical interferometry[1] as well as 3D reconstruction using fringe projection profilometry (FPP)[2], phase retrieval is a common but fundamental step. To achieve accurate and robust phase retrieval, various techniques and algorithms have been developed. Among state-of-the-art algorithms, PS algorithm[3] has been one of the most used algorithms due to its high performance. However, for a standard $N$-step PS algorithm, at least $N \geq 3$ fringe patterns with difference phase-shifts are needed, which limits the algorithm in a relatively narrow range of application. For instance, capturing multiple patterns in FPP makes the phase retrieval more error-prone to ambient disturbances, as well as limited to static scenarios. To reduce the number of fringe patterns, variants of PS have been present in previous works, such as $\pi$-shift PS[4]. Most recently, a two-step PS algorithm[5] for FPP was presented, where only two fringe patterns were captured. Based on variants above, the number of patterns is reduced to 2, without sacrificing performance significantly. Nevertheless, two-pattern PS algorithms are still time-consuming for some real-time scenarios. Furthermore, the mechanism used for realizing phase shifting makes the system more complicated and cumbersome.

For further improvement, single fringe pattern-based algorithms have been proposed. By exploiting spatial information of phases, phases are retrieved from only one fringe pattern using Fourier transform (FT) and its extensions, such as windowed Fourier transform, wavelet transform and S-transform, etc[6-10]. Even though extensive efforts have been made toward more accurate and robust phase retrieval, performances of these FT-based algorithms still significantly rely on hand-drafted filters and related parameters. Moreover, due to the non-pixel-wise operation, it is still challenging for FT-based algorithms to retrieve phases in surface discontinuity with sharp edges. To simulate PS and achieve pixel-wise phase retrieval, a digital 4-step PS algorithm was developed based on Riesz transform (RT)[11]. Given only one fringe pattern, three $\pi/2$ phase-shifted patterns can be generated using RT algorithm. With the resulting 4-step PS patterns, phases can be retrieved pixel-wisely using conventional PS algorithm. While RT algorithm performs well for patterns with high signal-to-noise ratio (SNR), the performance decreases dramatically especially for degenerated patterns by varying surface curvature and reflectance. Additionally, the generated PS step $N$ is limited to 4 due to the $\pi/2$ phase-shift, which makes the algorithm more sensitive to noises.

In this letter, we propose a deep learning model based digital PS algorithm, termed PSNet, which can accurately predict other PS patterns from only a single pattern. From a simulated dataset, the PSNet automatically learns digital N-step PS formation, and the adaptability to complex scenarios with varying surface curvature and reflectance. To the best of our knowledge, this is the first attempt to achieve digital N-step PS pattern generation from only one single pattern using deep learning model.

A typical captured fringe pattern with phase-shift can be expressed as

$$I_i(x, y) = a(x, y) + b(x, y)\cos\left[\varphi(x, y) + \delta(i-1)\right], \quad (1)$$

where $a(x,y)$, $b(x,y)$ and $\varphi(x,y)$ are average intensity, modulation and wrapped phase of the captured fringe pattern respectively. $(x,y)$ is the camera image coordinate pair. $\delta$ is the phase-shift, and $i=1,...,N$ indicates the pattern with $i$-th phase-shift and $N$ is the PS number.

With conventional PS algorithm, the wrapped phase $\varphi(x,y)$ can be retrieved using

$$\varphi(x, y) = -\arctan \frac{\sum_{i=1}^{N} I_i \sin(\delta i)}{\sum_{i=1}^{N} I_i \cos(\delta i)} . \qquad (2)$$

To generate $N$-1 patterns $I_i$, $i=2,...,N$, from a single pattern $I_i$, $i=1$, we designed the proposed PSNet. The model is constructed with a typical Unet[12] as the backbone. Detailed architecture of PSNet is shown in Fig. 1.

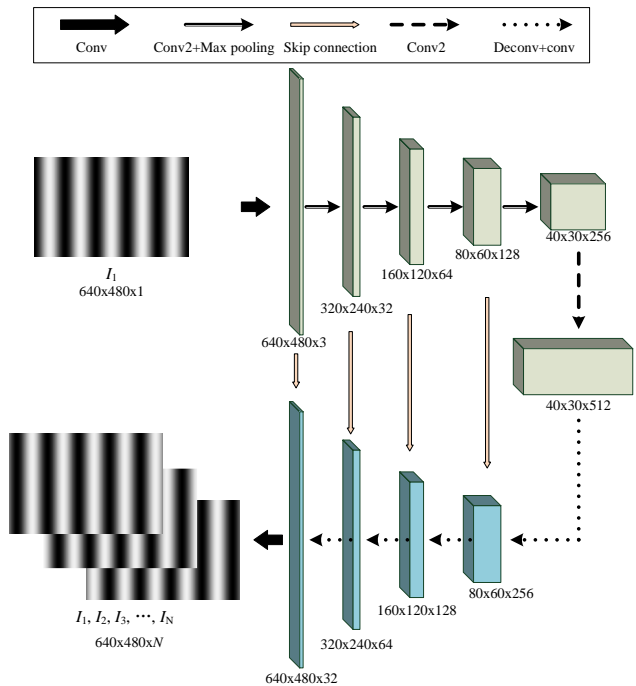

Fig. 1 Architecture of PSNet.

As shown in Fig. 1, the input of PSNet is a single fringe pattern with 640x480pixels, i.e. $I_1$, and the output is sequence of predicted N-step PS fringe patterns, i.e. $I_i$, $i=1,2,3,...,N$. please note that we include $I_1$ in the output, which can also be removed optionally. The PSNet has an approximatively symmetric architecture, as shown in Fig.1. Near the input end, there are one "Conv" layer, four "Conv2+Max pooling" layers and one "Conv2" layer, which forms the "down sampling" process and produces a 40x30x512 feature map. Conv indicates a typical convolutional layer. Conv2 means perform convolution operation twice sequentially on its input, and Conv2 followed by max pooling operation is the layer "Conv2+Max pooling". In contrast to the down sampling process, the up-sampling process includes four Deconv+conv layers and one Conv layer. Deconv+conv layer firstly performs deconvolution operation on the input, whose output is then concentrated with the Skip connection's output and processed with a convolutional layer. The sizes of all convolution kernels in is 3-by-3.

A simulated PS pattern dataset is also introduced for training the PSNet, which consists of roughly 180 objects with diverse shape, reflectance and poses. For each object in certain pose, a set of 8-step PS patterns were captured as one training sample in a simulation environment, where an FPP system of one 640x480pixel-camera and a 1280x800pixel-projector was implemented. Consequently, 5 samples were collected for the same object in 5 different poses, as shown in Fig. 2. Thus, the resulting dataset consists of 900 samples, where some examples are shown in Fig. 3.

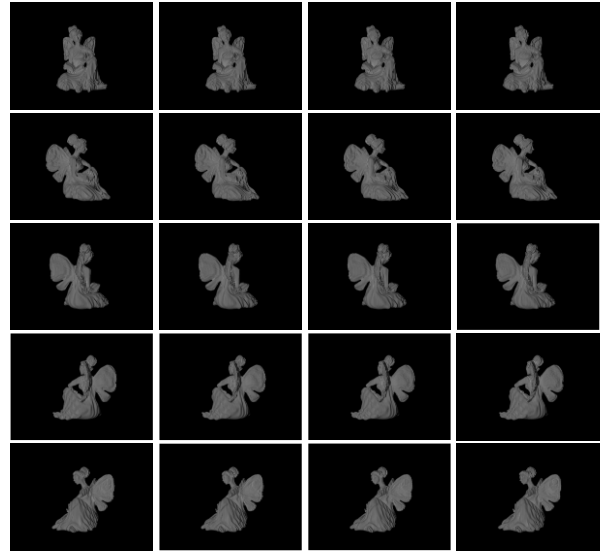

Fig. 2 One example of an object in the dataset, where each column represents patterns with different phase-shifts, i.e. $I_i$, i=1,3,5,7, and each row represents the object in different poses.

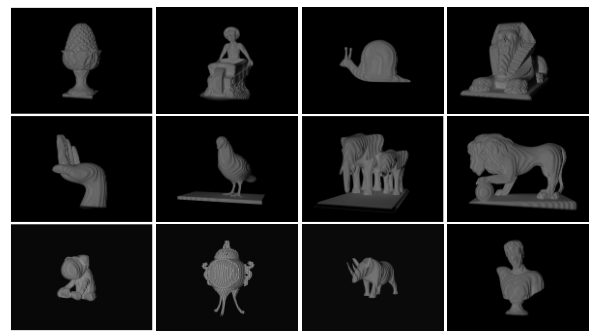

Fig. 3 Some samples in the dataset.

All the simulation environment was implemented in the platform of 3Ds Max 2018 [13]. Before training, the dataset was spilt into training, validation and testing sets, respectively, in a ratio of 8:1:1, resulting roughly 720, 90 and 90 samples in each set.

During the training of PSNet, some specific designs were used. Firstly, a training loss combining intensity and phase error is introduced, which is expressed

$$L = L_{\text{intensity}} + \lambda L_{\text{phase}}$$
$$= \sum_{i=1}^{N} \left\| I_i - I_i^* \right\|^2 + \lambda \left\| \varphi - \varphi^* \right\|^2 , \quad (3)$$

where $I_i^*$, $i=1,2,…,N$ are predicted PS patterns by PSNet, and $\varphi^*$ is the wrapped phase retrieved using predicted PS patterns according to Eq. 2. $\lambda$ is a weighting factor, which is set 0.015 empirically for best performance.

Secondly, considering image degeneration due to varying surface curvature, reflectance, and lighting conditions in the real scenario, data augmentation is used to improve the generalization of PSNet. Specifically, each sample is augmented randomly according to the following equation

$$I_{augmented} = kI_i + I_{\text{offset}} , \quad (4)$$

where $k$ and $I_{\text{offset}}$ are scale factor and intensity offset respectively. $k$ is selected randomly in the range [0.5, 1.2], and $I_{\text{offset}}$ is also randomly selected in a normalized range [-0.05, 0.05].

With these two specifically designed tricks, the prediction accuracy and adaptability to complex scenarios of PSNet are guaranteed. The training is implemented in Pytorch[14]. After about 7 epochs, the training performance tended to stop improving, and accordingly we stopped it in 20 epochs.

For the validation of the proposed algorithm, we compared the predicted PS patterns and corresponding retrieved phases with the real captured ones, where the latter is regarded as the ground truth(GT). Simulated and experiment results are shown as follows.

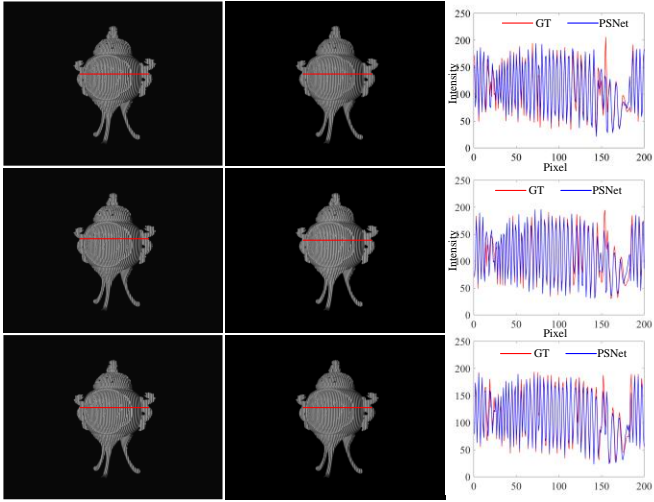

Fig. 4 PS patterns of a Tripod: from left to right, each column represents GT, predicted patterns with PSNet, and their cross-sections whose positions are marked in red lines in the first two columns; from top to bottom, each row presents fringe patterns with phase-shift $\pi/2$, $\pi$ and $3\pi/2$, respectively.

Simulation was performed based on the same FPP system implemented in the simulation environment. The FPP system in simulation consisted of a camera of 640x480 pixels and a projector of 1280x800 pixels. The fringe frequency and phase-shift step of projected fringe patterns are 1/10 and 4 step, respectively. Captured fringe patterns and the predicted ones are shown in Fig. 4. It is obvious that the predicted fringe patterns by PSNet are visually identical to the captured ones, i.e. GT. Furthermore, from a closer look at the cross-sections, the intensity distribution of predicted fringe is very close to GT, except for few regions with obvious error, as shown in Fig. 4. These results above suggest the promising performance of the propose PSNet on accurate prediction of PS patterns.

For further validation, additional comparison of retrieved phase was performed. As shown in Fig. 5, wrapped and unwrapped phases retrieved using predicted 4-step PS patterns are also very close to the GT. From the comparison of cross-sections in Fig. 5, only slightly small difference between the results of the proposed PSNet and GT can be found. Even for regions with obvious phase jumps in the cross-section of unwrapped phase, the proposed algorithm also achieved comparable performance, which provides evidence of the effectiveness of the proposed PSNet. Please note that the unwrapped phase was retrieved using multi-frequency fringe phase unwrapping (MFPU) algorithm[15], where the projected fringe frequencies for unwrapping were 1/80, 1/1280, respectively.

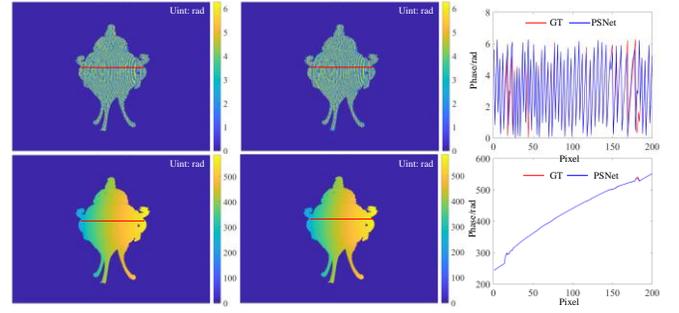

Fig. 5 Retrieved phases: from left to right, each column represents GT, result of PSNet and cross-sections, respectively, whose positions are marked in red lines in the former two columns; the top and bottom rows represent wrapped and unwrapped phases respectively.

To verify the generalization to PS patterns in real scenarios, additional real-scene experiments were conducted. An FPP system of a 3384x2704-pixel camera and 1280x800-pixel projector was established. Same frequency and phase-shift step patterns were projected on diverse objects by the projector and then captured by the camera. To adapt to the input format of PSNet, the captured patterns were resized from 3384x2704 to 640x480. Similarly, additional patterns for phase unwrapping were projected, where fringe frequencies were 1/80, 1/1280 respectively.

Results of two boxes are shown in Figs. 6 and 7. Similarly, no noticeable difference exists between the predicted patterns and GT, as shown in Fig. 6. Furthermore, comparable phase retrieval results were achieved using predicted patterns, as shown in Fig. 7, where only small phase errors can be found in phase error maps. To demonstrate the performance on discontinuous surface with large phase jumps, we compared cross-sections of phases marked in Fig. 7. As shown in Fig. 8, PSNet demonstrated comparable performance on wrapped and unwrapped phases respectively, comparing to the GT results. Particularly, around the discontinuity between the two boxes, there is an obvious phase jump, where the proposed PSNet performed as well as GT with only small deviations. These results above verified the proposed algorithm's effectiveness and generalization to real complex scenarios with obvious discontinuity.

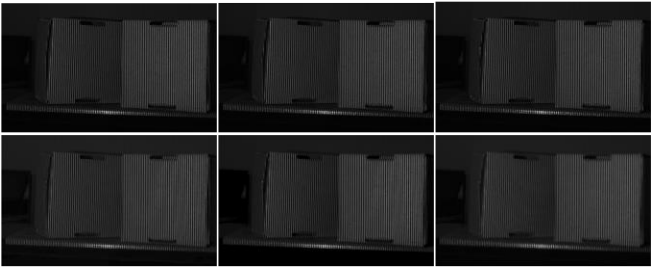

Fig. 6 PS patterns of boxes, from left to right, the top and bottom rows represent GT and predicted patterns with PSNet respectively, and each column represents patterns with phase-shift $\pi/2$, $\pi$ and $3\pi/2$ respectively, from left to right.

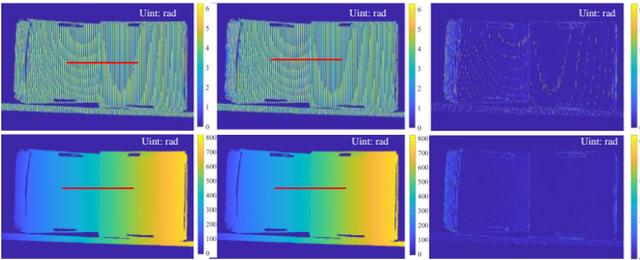

Fig. 7 Phases and errors: from left to right, each column represents GT, result of PSNet and phase error maps, respectively; the top and bottom rows represent wrapped and unwrapped phases respectively.

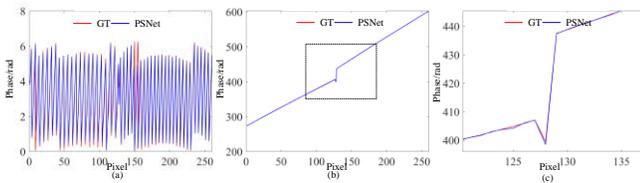

Fig. 8 Cross-sections: (a) and (b) are cross-sections of wrapped and unwrapped phases marked in red lines in Fig. 7, and (c) is the zoomed-in version of the curve boxed in dashed black line in (b).

To further demonstrate the capacity of arbitrary-step PS pattern generation, we tested the PSNet on 8-step PS patterns. As shown in Fig. 9, it is obvious that PSNet predicted PS patterns visually identical to the GT ones, and additionally, achieved accurate phase retrieval in contrast to the GT result.

In conclusion, a deep learning model based digital phase-shifting algorithm termed PSNet is proposed in this letter. Given only one single fringe pattern, PSNet can accurately predict patterns with other phase-shifts. Simulation result demonstrates the impressive performance on N-step PS pattern generation. Furthermore, PSNet generalizes well to real complex scenarios, and achieves comparable results on scenarios with discontinuous surface and large phase jumps. The proposed algorithm can be widely used in phase retrieval in optical interferometry and FPP, which can dramatically reduce the time costing on captured multi-patterns.

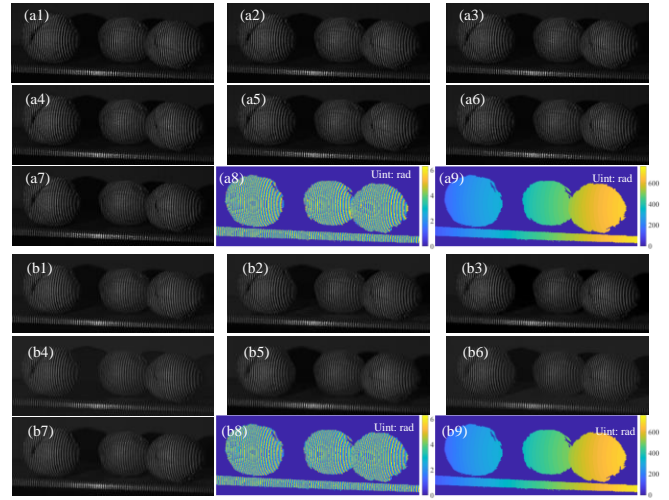

Fig. 9 Eight-step PS patterns and retrieved phases: (a1)-(a7) are GT patterns with phase-shifts $\pi/4$, $\pi/2$, $3\pi/4$, $\pi$, $5\pi/4$, $3\pi/2$, and $7\pi/4$ respectively, and (b1)-(b7) are predicted ones; (a8)-(a9) are wrapped and unwrapped phases, and (b8)-(b9) are ones of PSNet.

**Funding.** National Natural Science Foundation of China (NSFC) (62005217); Fundamental Research Funds for the Central Universities (No. 3102020QD1002); the 111 Project (B16039).

**Disclosures**. The authors declare no conflicts of interest.

## Full References